%% file: doc.tex
\documentclass[10pt]{elsarticle}

\usepackage{hyperref}

\input{revision}

\ifcsname draft\endcsname%
\usepackage{todonotes}
\else
\usepackage[disable]{todonotes}
\fi

\ifcsname anoniem\endcsname%
\newcommand{\anonimize}[2]{[#2]}
\author{}
\address{\small revision \revision}
\else
\newcommand{\anonimize}[2]{#1}
\author{Wauter Bosma, Sander Dalm, Erwin van Eijk, Rachid el Harchaoui, Edwin Rijgersberg, Hannah Tereza Tops, Alle Veenstra, Rolf Ypma}
\ead{w.bosma@nfi.nl}
\ifcsname draft\endcsname%
\address{\small revision \revision}
\else
\address{doi: \href{https://doi.org/10.1016/j.scijus.2019.10.005}{10.1016/j.scijus.2019.10.005}\\
© 2020. This manuscript version is made available under the CC-BY-NC-ND 4.0 license http://creativecommons.org/licenses/by-nc-nd/4.0/}
\fi
\fi

\title{Establishing phone-pair co-usage by comparing mobility patterns}
\date{\small rev. {\revision}}

\begin{document}

\begin{abstract}
In forensic investigations it is often of value to establish whether two phones were used by the same person during a given time period. 
We present a method that uses time and location of cell tower registrations of mobile phones to assess the strength of evidence that any pair of phones were used by the same person.
The method is transparent as it uses logistic regression to discriminate between the hypotheses of same and different user, and a standard kernel density estimation to quantify the weight of evidence in terms of a likelihood ratio.
We further add to previous theoretical work by training and validating our method on real world data, paving the way for application in practice.
The method shows good performance under different modeling choices and robustness under lower quantity or quality of data.
We discuss practical usage in court.
\todo{labels op alle assen vd plaatjes}
\end{abstract}


\maketitle

\section{Introduction}

As mobile phones have become established in everyday communication, so has their use in criminal activity.
Criminals often use cheap, prepaid ``burner phones'' to communicate about criminal activities. These burner phones are often used for some time before being discarded. 
Users of burner phones often carry multiple phones at the same time, including a legitimate, ``personal'' mobile phone to contact family and friends.

Identifying the user of a burner phone can be of great value in tying criminal activities to suspects.
One type of traces that may reveal the user's identity is location traces, as derived from call detail records (CDRs), by comparing the location data from the burner phone to location data from the personal phone of a suspect.
CDRs are stored by telecom providers and can typically be requested by police for a period of up to several months in the past, even if the phone was never found.

Traditionally, forensic analysis of location data is centered around key events, and focused on few individual records from CDR files.
However, CDRs suffer from low accuracy when it comes to pinpointing the position of a cell phone user.
While individual records may show whether or not a cell phone is within a certain area, they are less suitable to determine whether the phone is with the same user as another phone.
On the other hand, when CDRs are available for an extended period of time, there is no need to restrict this comparison to one or several records.
A series of CDRs of two phones can provide statistical evidence that the two phones were carried by the same user.

In this paper, we propose a method of comparing mobility patterns from CDRs of two phones, using logistic regression.
The outcome of this process is a similarity score which indicates whether the phones traveled with the same or with a different user.
To attain usability in court, we use this similarity score to calculate evidential strength in terms of a {\em likelihood ratio}.
The likelihood ratio expresses how probable the similarity score is in the scenario that both phones belong to the same user, relative to the scenario that both phones belong to different users.
We performed field experiments to obtain the reference set needed for this approach. 
Furthermore, we tested the method's performance on data from real phone usage, which was not used to train the model.
Finally, we evaluated the robustness of this method by evaluating performance under a range of different scenarios and method perturbations, including a comparison to a simpler approach based on the notion of {\em dislocations}.

\section{Background and related work}

When a mobile phone communicates, be it for calling, texting or data traffic, it connects to a cell tower. The timestamp of this event, as well as the identification number (cell id) of the tower are recorded by the telecom provider in call detail records.
The location of this cell tower gives an approximation of the location of the mobile phone user, because a mobile phone will preferentially connect to the cell tower that it has the best connection with in terms of signal strength.
This means mobile phones are more likely to connect to nearby cell towers. Still, many factors may influence which cell tower a phone connects to, including:
\begin{itemize}
 \item the transmission frequency;
 \item the density of cell towers in an area;
 \item the transmission power of the cell tower and phone;
 \item obstacles interfering with line of sight;
 \item network congestion; and
 \item weather conditions.
\end{itemize}

The location of a phone user can be estimated from call detail records. The uncertainty of this estimate varies from several kilometers in rural areas with low cell tower density, to mere hundreds of meters in urban areas \cite{vanbree15a}.

\subsection{Analysis of mobile phone activity patterns}
Analysis of mobile phone usage patterns is an area that has seen previous study. A notable finding by de Montjoye et al. \cite{demontjoye13a} is that the spatiotemporal patterns that emerge as a result of a mobile phone's activity can be used to identify its user. With four spatiotemporal data points, the pattern is unique for as much as 95 percent of users.
The more fine-grained the spatial and temporal information, the more accurately a user can be identified.
It is even possible to predict where a cell phone user will go next, given his/her historical movement patterns \cite{lu13a, ozer14}.
Mobile phone communication patterns can also be used to determine social closeness between phone users \cite{phik10a,phik11a}. 

Looking at a forensic context, the traditional use case for cell site analysis is to determine whether or not a certain individual could plausibly have been at a crime scene at a certain time \cite{vanbree15a}.
For this purpose, only a small number of observations are of interest, namely those surrounding the moment of the crime.
In the present research, by contrast, the spatiotemporal pattern of a pair of mobile phones is analyzed over an arbitrary length of time in order to determine their similarity. 

Our research builds on the pioneering work by Van Eijk \cite{vaneijk13a}. 
The author explored the concept of a \textit{dislocation} for the purpose of determining whether two phones are moving together. 
A dislocation is the observation that two phones are used close together in time, but at a large geographic distance, meaning that a single person could not possibly be responsible for their usage. 
Based on a simulated dataset, Van Eijk laid out some characteristics of dislocations and their usefulness in identifying phone users.
However, it is of yet unknown whether the concept of dislocations is sufficient or optimal in the presence of real-world data. 
Another reason for looking beyond dislocations is that we
do not only wish to rule out individuals as users of a phone, but also provide a statistical estimate of the similarity of two mobility patterns.
For example, if a population of phone users all live in the same geographical area, not many dislocations may occur,
but still some mobility patterns may be more similar than others. This is something we wish to quantify.

In the present research, real world data are used and machine learning methodology is applied to determine the similarity of mobility patterns. Then, a likelihood ratio is calculated to estimate the probability of the evidence
given the hypothesis of the phones having the same user and the hypothesis of the phones having different users.

\subsection{Calculating likelihood ratios in a forensic context}
In our method, the mobility patterns extracted from call detail records are used to calculate the strength of evidence, expressed as a likelihood ratio (LR) of two phones traveling together.
LRs and Bayesian reasoning have been used in criminal cases for several decades \cite{finkelstein70a}. Still, the question to what extent statistics and Bayesian reasoning have a place in the court room is sometimes debated \cite{tribe71a}.
In 2010, a UK Court of Appeal ruled that Bayes' theorem and LRs  should not be used, other than for the quantification of DNA evidence. It was argued that the use of likelihood ratios for non-DNA evidence was not `statistically sound' \cite{ukcourt10a}.
Fenton et al. strongly counter the position of the UK Court, stating that it was based on a poor understanding of Bayesian logic \cite{fenton14a, fenton14b}. Mainly, they point out that no database is perfectly reliable and that it is a misconception that
DNA databases are reliable \textit{because} they yield high LRs. In fact, they argue, there is no reason to declare DNA databases `more scientific' than other forensic databases.

Returning to the present study, given the inherently probabilistic nature of location patterns extracted from call detail records, we argue that expressing our findings in terms of an LR is the best way of quantifying evidence based on mobile phone traces.
Reporting on the similarity of two mobile phone traces without reporting on the probability of a random match gives an incomplete picture of the nature of the evidence.
In a forensic context, the LR is defined as:

\begin{equation}
	\frac{\Pr{(E|H_p)}}{\Pr{(E|H_d)}},
\end{equation}

where $\Pr{(E|H_p)}$ refers to the probability of the evidence given the prosecutor's hypothesis, and $\Pr{(E|H_d)}$ to the probability of the evidence given the defense hypothesis.
In this paper, it is forensically relevant whether two telephones had the same user during a time period. Therefore, our LR is defined as:

\begin{equation}
\frac{\Pr{(E|H_{su})}}{\Pr{(E|H_{du})}},
\end{equation}

with $\Pr{(E|H_{su})}$ the probability of the evidence given that the telephones had the same user and $\Pr{(E|H_{du})}$ the probability of the evidence given that the telephones had different users.

\section{Materials and methods}

\begin{figure}
\centering
\includegraphics{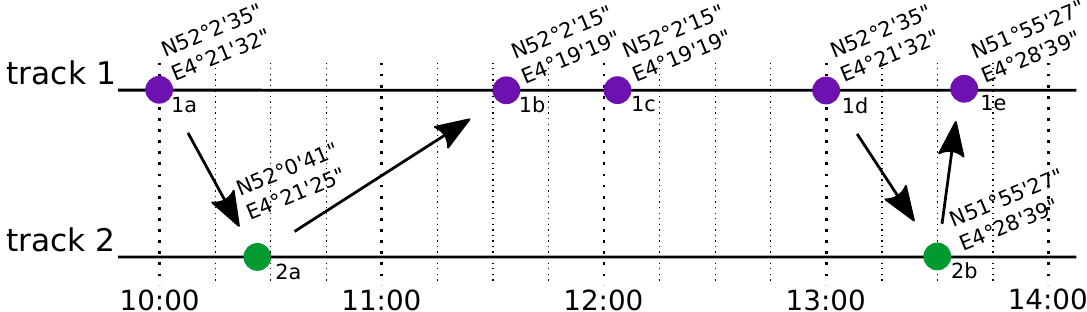}
\caption{A pair of cell phone location tracks on a timeline.
The dots represent measurements (derived from call detail records), each with geographic coordinates and a timestamp.
Arrows indicate consecutive measurements from both phones, which we refer to as \textit{switches}.
This particular pair of tracks has four switches: 1a+2a, 2a+1b, 1d+2b, 2b+1e.
\label{fig:tracks}}
\end{figure}

We will assume that two mobile phones $m_1$, $m_2$ that have the same user will be in the same place at all times.
The call detail records we use to determine the location of a mobile phone are only generated at certain events, such as the initiation of a phone call.
This means we will have a limited number of observations for each phone during a period of comparison, where an observation consists of a time $t$ and the location of the cell tower $x$.
The data for each phone $m_i$ thus consist of a vector of time-location measurements, $\{t_j^{m_i}, x_j^{m_i}\}_{j=1,...,n}$.
We split the data for each phone into, potentially multiple, periods of 15 hours between 7 AM and 10 PM.
We will refer to the vector of data points for a certain phone within one period as a {\em track}.
Figure \ref{fig:tracks} shows the measurements extracted for a pair of tracks.

\begin{figure}
\centering
\includegraphics[width=.5\textwidth]{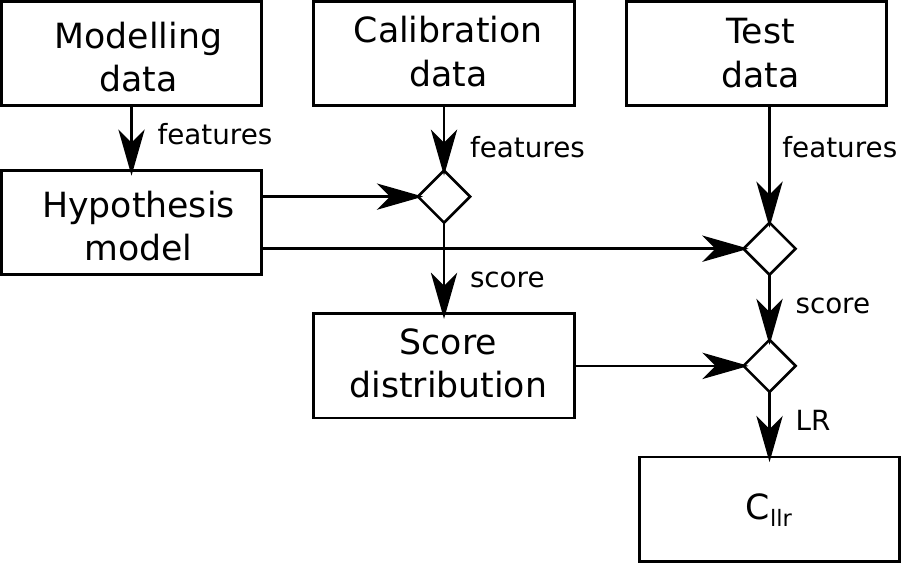}
\caption{The experimental setup: training data is used for constructing a model which calculates hypothesis scores for phone pairs.
Calibration data is used to infer the probability distributions of scores for both hypotheses.
Test data is used to evaluate the system, calculating performance metrics such as $C_{llr}$.
\label{fig:opstelling}}
\end{figure}

Our objective is to measure similarity of a pair of tracks, where a high degree of similarity supports the hypothesis of the phones having the {\em same user}, and a low degree of similarity supports the {\em different user} scenario.
Our experiments are outlined in Figure \ref{fig:opstelling}.
It consists of the following steps.
We collect data for both scenarios ($H_{su}$ and $H_{du}$), as described in section \ref{sec:datacollection}.
We randomly split the data into a 80 percent training set (40 percent for model construction and 40 percent calibration) and a 20 percent test set (for evaluating performance).
For each track pair in the test set, we calculate the similarity (section \ref{sec:transition-classifier} and \ref{sec:track-classifier}).
From the similarity score, we compute an LR using a kernel based approach (section \ref{sec:LR-computation}).
Finally, we analyze the performance and characteristics of our method (section \ref{sec: analysing-method}).

\subsection{Data collection}
\label{sec:datacollection}

Our method requires samples of {\em same-user} track pairs and {\em different-user} track pairs.
The same-user track pairs should be sampled from a pair of phones which belong to the same user.
The different-user track pairs should be sampled from pairs of distinct users, preferably belonging to the same population of interest.
In real-world applications, the characteristics of these same-user and different-user pairs should resemble the characteristics of the research phones in question.
In our experiment, the data set is characterized by the specific population of phone users.

In theory, measurements of a phone pair could be sampled from a single phone, where each measurement is assigned randomly to either of two tracks.
However, this would ignore the fact that two phones in the same location may use different cell towers, even if they use the same telecom provider.
We therefore conducted field experiments (see below) where individual subjects carried around three telephones registered to three different networks.
We used these data to develop and test our method. We further acquired an anonimized set of call detail records from real cell phone users to validate the method.

\subsubsection{Field experiments}

We conducted field experiments using specifically prepared phones.
These phones were carried by 18 test subjects (\anonimize{NFI}{ORGANIZATION} employees) for at least one week each, to ensure we obtain location data from natural behavior.
Each subject carried a package of three phones, generating same-user mobility patterns.

All three major telecom providers were used --- each of the phones was registered to one of these.
The prepared phones had the open source software \textit{NetworkMonitor}\footnote{\tt https://github.com/caarmen/network-monitor} installed on them,
which logs the cell tower they are currently connected to every 10 seconds.
So, all three phones within a set were with the same person at the same location at all times.
At times, a phone would briefly run out of power, meaning that we did not always collect a full export for all three phones in a set.
In total, we collected data for 17806 hours (on average 14 days for 18 users, carrying 3 phones).

\subsubsection{Synthesizing CDRs from experimental data}
There are two aspects of the space-timestamps provided by CDRs that are informative of possible co-usage of a pair of phones.
First, the geographical trajectories, e.g. whether the phones appeared to be close to each other at all times.
Second, the temporal pattern of events, e.g. the correlation in call times for the two phones.
Using the described data collection method we can mimic the former, but not the latter, as there was no usage of phones, simply a log entry every 10 seconds.
This limitation is partly for practical reasons, we cannot `force' test subjects to naturally use an extra phone, and partly for theoretical reasons.
We strongly doubt that temporal usage patterns of the test population would be representative for the usage patterns of a criminal with a `private' and burner phone.
It may even be opposite --- correlation in the field experiments (a call on your private phone triggering a call on your work phone) versus anti-correlation in criminal investigations (intentional use of both phones on different parts of the day).
In fact, we neither know, nor have the data to estimate this.
Thus, we take the conservative approach of sampling usage times independently for each phone. 
This way, there is no information in just the temporal distances for the algorithm to learn from.
This makes our approach conservative in the sense that 1) we avoid learning to predict co-usage based on characteristics of temporal features that may not hold for the population of interest and 2) LR estimations will tend be closer to 1 than if we had used both aspects above, as less information is available.

\begin{figure}
\centering
\hspace*{-10mm}%
\includegraphics{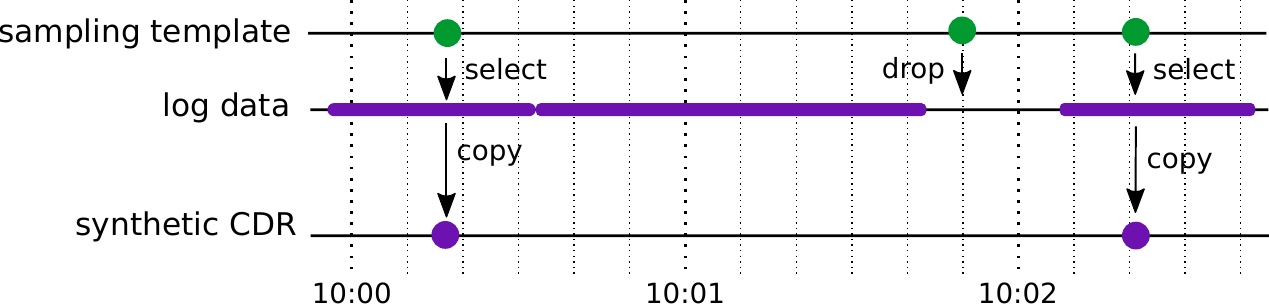}
\caption{Synthetic call detail records (CDRs) are generated from NetworkMonitor log data and real CDRs. For each real CDR entry the NetworkMonitor log entry closest in time is selected. The synthetic CDRs consist of the set of selected log entries, and thus simulate natural temporal usage patterns. Pairing synthetic CDRs generated from co-used phones and different real CDRs yields same-user track pairs.
\label{fig:synthesis}}
\end{figure}

Concretely, we use a very simple model of human call behavior, simulating usage patterns by sampling event times from a homogeneous Poisson process with a rate of one per hour.
These sampled time points were then assigned the coordinates from the log entry closest in time, as illustrated in Figure \ref{fig:synthesis}.
The result is a synthesized CDR file which to some extent mimics natural behavior, where we chose a simplified model for the reasons stated above.

We now have synthesized CDRs of variable length. We extract 15 hour tracks from these, by taking the 7 AM - 10 PM period for each day.
In this time frame, users are generally active, both moving around and making calls. This procedure also ensures all tracks
are of the same length. Tracks which covered less than 80\% of the time period were discarded, as well as tracks that showed little movement ($<$ 10 km summed over all consecutive events).

To form the track pairs, we only matched tracks from the same date. This ensures we don't compare week days to weekend days,
for which we could expect quite different behavior.
We refer to these data as reference data.

\subsubsection{Real user data for validation}

To validate our use of synthesized data, we acquired an anonimized set of call detail records from 22 real cell phone users who volunteered to share their data.
This data set contains two phones for 15 of these users, typically one for work and one for private use.

The users are members of the same social network, and live in the same region.
This is a suitable population for our purposes, as in court cases the possible other users of a burner phone will often be from the social network of the defendant.
Furthermore, as the location of two of the real cell phone users may incidentally coincide,
distinguishing between the two hypotheses will be harder for these two users than if they had been selected from a large population at random.

The call detail records were provided by the major telecom providers operating within \anonimize{the Netherlands}{GEOGRAPHIC AREA}.
All providers produced call detail records including the cell id, but in some cases also the geographic coordinates of the cell tower were provided.
Where necessary, we used an external database, provided by the providers, to map cell towers to geographical locations. 
Again, tracks which covered less than 80\% of the time period were discarded, as well as tracks that showed little movement ($<$ 10 km summed over all consecutive events).

\begin{figure}
\centering
\includegraphics[width=.8\textwidth]{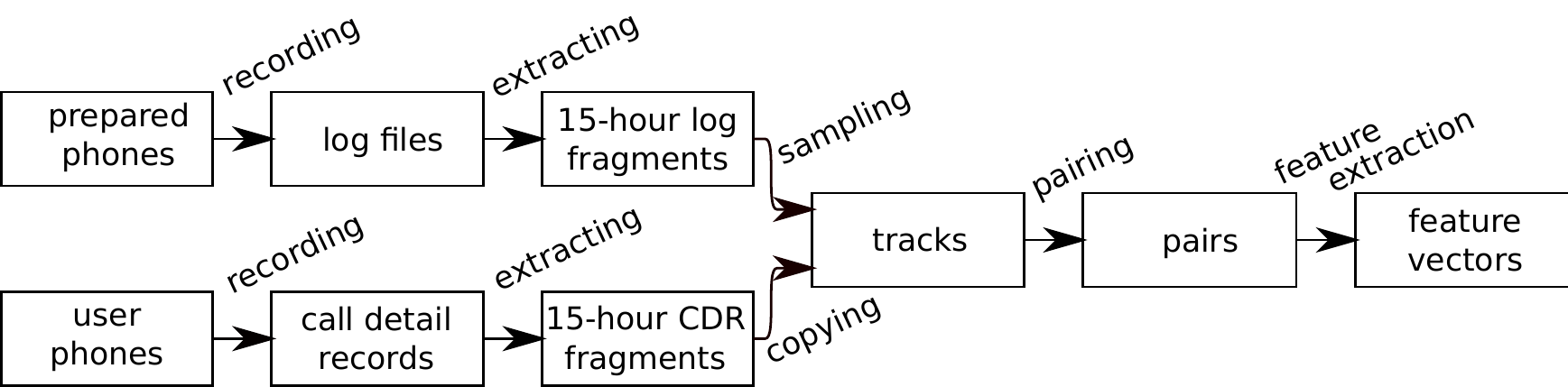}
\caption{The preprocessing pipelines of both types of data produce 15-hour tracks, which are paired into same-user and different-user track pairs. However, the user phones generate CDRs, while the prepared phones produce continuous log files which are subsampled to mimic CDRs.
\label{fig:pipeline}}
\end{figure}

Figure \ref{fig:pipeline} outlines the full preprocessing pipeline. 
Note that the real user data are not used to fit or calibrate the model, but only to assess its performance.
We refer to these data as validation data.

\subsection{Scoring adjacent measurement pairs}
\label{sec:transition-classifier}

Our goal is to determine whether two mobile phones $m_1, m_2$ share a user or have distinct users.
Intuitively, the most informative measurements are those where both devices are used within a short time period.
If those measurements are geographically distant, it is more likely that the devices in the pair belong to different users.
Therefore, as a first step, we identify measurement pairs originating from both tracks which are adjacent in time.
We will follow \cite{vaneijk13a} in referring to these pairs as switches.
That is, a measurement pair $[(t_k^{m_i}, x_k^{m_i}),(t_l^{m_j}, x_l^{m_j})]$, with $t_k^{m_i}$ the time phone $m_i$ connects with a cell tower having location $x_k^{m_i}$, is selected and called a switch if and only if $i\neq j$, and $t_k^{m_i}<t_l^{m_j}$, and there exists no $t_n^{m_o}, o\in\{i,j\}$ such that $t_k^{m_i} < t_n^{m_o} < t_l^{m_j}$ (see also Figure \ref{fig:tracks}).
For each of these switches, we derive a set of features, and we use those features to fit a model which outputs a score between 0 and 1, with higher scores indicating that the switch belongs to the {\em same-user}.

The following features are computed for each switch:

\begin{enumerate}
    \item the distance $d(x_k^{m_i}, x_l^{m_j})$, as the crow flies, between the geographical locations of the two cell towers in meters;
    \item the time difference $t_l^{m_j}-t_k^{m_i}$ between each of the registrations in seconds;
    \item the `speed', i.e. the distance divided by the time difference $\frac{d(x_k^{m_i}, x_l^{m_j})}{t_l^{m_j}-t_k^{m_i}}$, in meters per second.
\end{enumerate}

The distributions of the different features have very different mean values.
Because statistical models often perform better with more evenly distributed features, we rescale the features (subtracting the median and scaling by the interquartile range).
A logistic regression model was used to model whether a switch was from a {\em same-user} or {\em different-user} track pair.
The result is a score between 0 and 1, for each switch.

The choice for logistic regression was based on the following considerations:
\begin{enumerate}
    \item logistic regression is a relatively simple and linear model, which improves overall model generalization;
    \item the model provides coefficients for each feature. In this respect the model is not a black box, but a transparent system that can be explained to lay people, such as magistrates.
\end{enumerate}

The end result is a score between 0 and 1, for each switch.

\subsection{Scoring track pairs}
\label{sec:track-classifier}

To compute a similarity score for a pair of tracks, the scores of the individual switches need to be aggregated into a single score.
We bin the switch scores into ten bins with a range of 0.1 points each, and normalize these counts to arrive at ten fractions together summing to 1.
These ten fractions then serve as the features for a second logistic model that predicts whether a pair of tracks has the same or a distinct user.

Thus, the two logistic regression models together compute a similarity score between 0 and 1 for any pair of tracks, where 1 expresses
strong support for $H_{su}$ and 0 for $H_{du}$.

\subsection{Computing the likelihood ratio}
\label{sec:LR-computation}

\begin{figure}
\centering
\includegraphics[width=.8\textwidth]{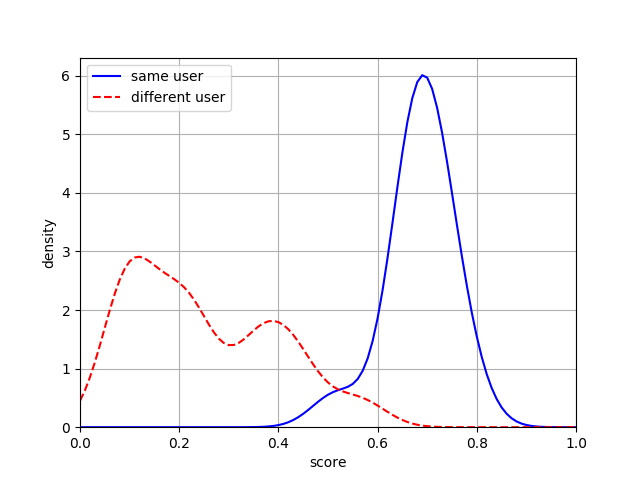}
\caption{ Probability distribution of track-pair scores for both hypotheses.
\label{fig:score-density}}
\end{figure}

We sample from the probability density of a score $s$ given a hypothesis $H_i$, by computing scores for all track pairs for which $H_i$ holds.
We can thus obtain the empirical probability density function.
We used Kernel Density Estimation (KDE) with a Gaussian kernel, with bandwidth chosen by visual inspection as 0.05, as shown in Figure \ref{fig:score-density}.
The LR is the ratio of the two probability density functions $\frac{\Pr{(s|H_{su})}}{\Pr{(s|H_{du})}}$.
Finally, we note that determination of the most extreme LRs is sensitive to arbitrary choices that govern estimation of the tails of the distributions of scores \cite{elub}.
We therefore bound the computed LRs using a method motivated by information theory \cite{elub}, which demands that an LR system outperforms the neutral system of always giving back an LR of 1, even in the presence of a misleading data point (e.g. a large LR when $H_d$ is true). Bounding the generated LRs using this method ensures we are conservative in extrapolating given our data set size.

\section{Analysing the method}
\label{sec: analysing-method}
We refer to the method as described in sections \ref{sec:transition-classifier}-\ref{sec:LR-computation} as the baseline method.
We performed a range of sensitivity analyses to assess how performance of this baseline method alters when
\begin{enumerate}
\item a simpler approach than the baseline two-step method is used (section \ref{one-step})
\item other classifiers than logistic regression are used (section \ref{classifiers})
\item other calibration methods are used (section \ref{calibrators})
\item the percentage of train data used for model construction vs model calibration is altered (section \ref{calibration_fraction})
\item a subset of features are used (section \ref{features})
\item data are reduced in quantity (section \ref{robustness})
\end{enumerate}

For each of these scenarios, we randomly split the data into 80\% train data and 20\% test data 100 times, and report on the average metrics found, as well as their standard deviations.
To assess whether the results generalize to real data, for each scenario we also report on results on the validation data.
\subsection{Using simpler models}
\label{one-step}
As described earlier, our aim is to assess pairs of tracks.
Each pair of tracks consists of a (potentially different) number of switches.
The baseline method employs two classifiers: the first classifier learns a score for each pair of log entries,
the second classifier takes a binning of these scores as features to classify the track pairs.
Although this is an intuitive approach, simpler methods, e.g. employing only one classifier, may well perform at a similar level.
As simpler models should be preferred, all other things being equal, we investigate the performance of three simpler models.

First, we compute a one-dimensional binning of the switches in the following way. 
For each feature, a linear binning with 10 bins is constructed using the minimum and maximum feature values found in the training data. 
The normalized counts are then computed, where any measurements in the test outside of the extremes are put in the minimum/maximum bin. 
This results in 10 values per feature, which sum to 1. 
Finally, these feature specific values are concatenated, resulting in a feature vector with size 30. 
This vector is used for classification of the track pair. 
Note that this method ignores the dependence between the features.

Second, we explore the intuition that to exclude the hypothesis of two tracks having derived from the same user, we need only a single switch that is inconsistent with this hypothesis.
This is the notion of {\em dislocation} introduced by \cite{vaneijk13a}.
Following their definition, for every track pair, we compute the number of switches for which the time difference is less than 15 minutes, and the distance is larger than 25 kilometers.
We then train a logistic regression classifier on this feature to compute track pair scores.

Third, again suggested by Van Eijk \cite{vaneijk13a}, we compute the mean distance over all switches for each track pair, and use that as the sole feature.

\subsection{Using other classifiers}
\label{classifiers}
The baseline method uses logistic regression to assign a score to each switch, and again logistic regression to assign a score to each track pair (based on their switch scores).
Logistic regression is a relatively simple method that we selected mainly for its explainability.
It is however likely that more sophisticated, non-linear models would give better performance.
We therefore performed experiments where we used well-known models from the machine learning literature.
In particular, we performed three experiments were both the model trained to score switches and the model trained to score track pairs were a) a support vector machine \cite{svm} b) a random forest \cite{randomforest} or c) gradient boosting \cite{gradientboosting}, using the implementations provided by the Python package {\it sklearn} with default parameters.

\subsection{Using other calibration methods}
\label{calibrators}
The baseline method uses kernel density estimation (KDE) with a manually selected bandwith of 0.05 to derive LRs from the scores provided by the classifier.
We performed additional experiments using a) no calibration, b) KDE with higher (0.1) or lower (0.025) bandwith, c) a fitted Gaussian rather than KDE and d) isotonic regression (the PAV algorithm).

\subsection{Allocating data to model training or calibration}
\label{calibration_fraction}
The baseline method uses half of all training data to fit the logistic regression models, and the other half for calibrating the resulting scores.
It is however not clear whether this allocation is optimal in any way.
As we have been unable to find much guidance in the literature,
we also present experiments where the fraction of training data used for calibration ranges from 0.05 to 0.95.

\subsection{Changing features}
\label{features}
The baseline method used speed, distance and time as features. These are (non-linearly) dependent. 
Using fewer features could negatively or positively impact results.
We therefore investigated the impact of dropping each of the three features used consecutively, or using each independently.
Having only one feature drastically reduces information, we thus expect the method to perform worse.

\subsection{Impact of data quantity and quality}
\label{robustness}
We performed experiments with differing quantity and quality of data for two reasons.
First, we expect method performance to correlate with data set size.
Thus, this section offers a validation of the method. 
Second, it is of interest what the minimal data requirements are to expect meaningful LR estimations.
We performed two sets of experiments.
First, we increased data quantity by taking tracks that lasted multiple, rather than only one day. To ensure constant data quality, we again only took the switches resulting from events between 7 AM and 10 PM.
Second, we performed experiments where we decreased the hourly rate of events, leading to events more distant in time. 
We concurrently increased the number of days to ensure the same expected number of events and switches for these experiments.

\subsection{Measuring performance}
\label{sec:performance}
The final outcome of our method is a likelihood ratio. Not only does this number entail a binary classification, with $\textrm{LR}>1$ implying evidence for $H_{su}$, the value of the LR also gives a measure of how strong the evidence is. Thus, a method should be considered better both when its classifications are correct more often, and when it rates evidence for correct classifications as strong but evidence for incorrect classifications as weak.

One metric that takes both these requirements into account is the log likelihood ratio cost ($C_{llr}$). This metric stems from information theory \cite{bruemmer_preez2006} and is recommended for use in evaluating LR-yielding methods \cite{meuwly2016}. Given a set of observations $O_{su}$ sampled under $H_{su}$, and $O_{du}$ sampled under $H_{du}$, with cardinality $N_{su}$ and $N_{du}$, the $C_{llr}$ for the method under scrutiny $M$ is defined as
$$
C_{llr,M}(O_{su}, O_{du}) = \frac{1}{2}\left(\frac{1}{N_{su}}\sum_{o\in O_{su}}log_2(1+\frac{1}{LR_M(o)}) + \frac{1}{N_{du}}\sum_{o\in O_{du}} log_2(1+LR_M(o)) \right),
$$
where $LR_M(o)$ indicates the LR yielded by the method on observation $o$. Note that the $C_{llr}$ is a cost, so a higher number indicates worse performance. A perfect method would yield a score of 0, whereas a method that never finds evidence, i.e. assigns an LR of 1 to every observation, would yield a score of 1.

A derived metric is the $C_{llr,min}$.
This is the $C_{llr}$ calculated after applying the Pool Adjacent Violators (PAV) transformation  to the set of LRs \cite{bruemmer_preez2006}.
This transformation can be seen as a form of post-calibration, resulting in a perfect calibration on the evaluation data.
The $C_{llr,min}$ is a metric of the discriminatory value of the LR system, and the difference $C_{llr} - C_{llr,min}$ is the calibration loss, or $C_{llr,cal}$.

We measure performance both on the test fraction of the reference data (`test data'), whose distribution corresponds to the train data, and on the real user validation data.
We expect performance to be always better on the first set.
Performance on the validation data allows us to assess generalizability, and may be more indicative of performance in criminal cases.

\section{Results}

The 18 independent field experiments resulted in 273 {\em same-user} track pairs
 and 428 {\em different-user} track pairs, with a total of 3847 and 6044 switches, respectively.
The real user data resulted in a validation dataset of 28 {\em same-user} and 300 {\em different-user} track pairs,
with a total of 844 and 4750 switches, respectively.

\begin{figure}
\hspace*{-7mm}%
\rotatebox[origin=c]{90}{\em user phones \hspace*{32mm} reference phones}~~%
\parbox{\textwidth}{
\centering
\em same-user \hspace*{4cm} different-user \\
\includegraphics[width=0.49\textwidth]{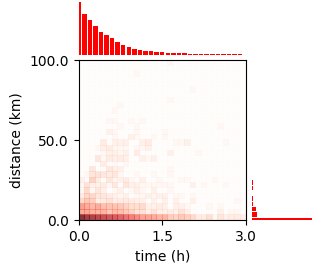}%
\includegraphics[width=0.49\textwidth]{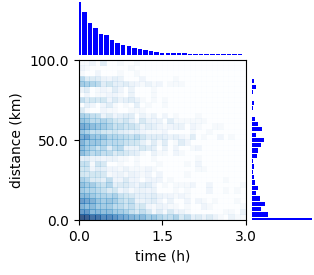} \\
\includegraphics[width=0.49\textwidth]{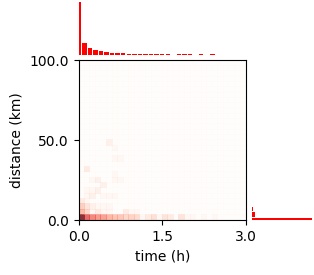}%
\includegraphics[width=0.49\textwidth]{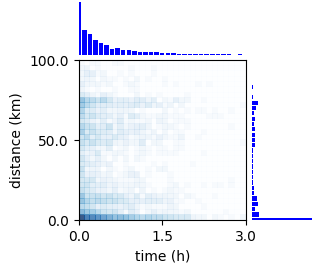} \\
}
\caption{
Heatmap and marginal distributions of geographical distance and time difference for all switches, for (left) {\em same-user} and (right) {\em different-user} track pairs,
for (top) the {\em reference phones} and (bottom) the {\em user phones}.
}
\label{fig:data}
\end{figure}

Figure \ref{fig:data} gives an overview of the distribution of features.
As expected, for the reference phones the time difference distribution is nearly identical for the two hypotheses, as the event times are sampled independently for phone pairs.
For the real user phones, the time difference distribution does differ for {\em same-user} and {\em different-user} pairs.
For both phone types, a clear difference exists in the distribution of distances between {\em same-user} and {\em different-user} pairs.
Likewise, both phone types have a tail in the distance distribution of the {\em different user} pairs that is non-decreasing, with higher peaks around 50 km for the reference phones, and around 75 km for the real user phones.
This is likely a result of the typical distance between the residences of the persons carrying the phones.
Increasing the sample size should lead to a flattening out of these tail, leading to a distribution of typical distances between residences of random people in the population.

\begin{figure}
\centering\footnotesize
\hspace*{5mm}test data \hspace*{45mm} validation data \\
\includegraphics[width=0.49\textwidth]{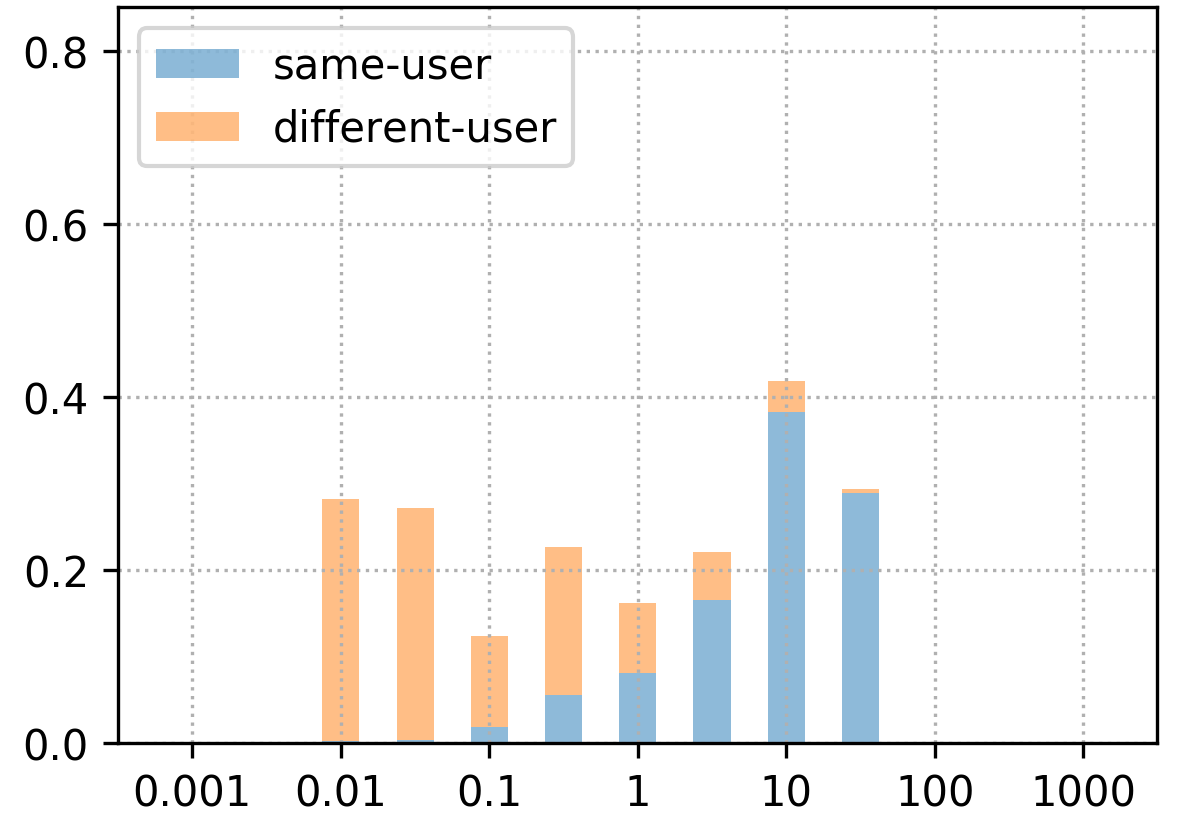}%
\includegraphics[width=0.49\textwidth]{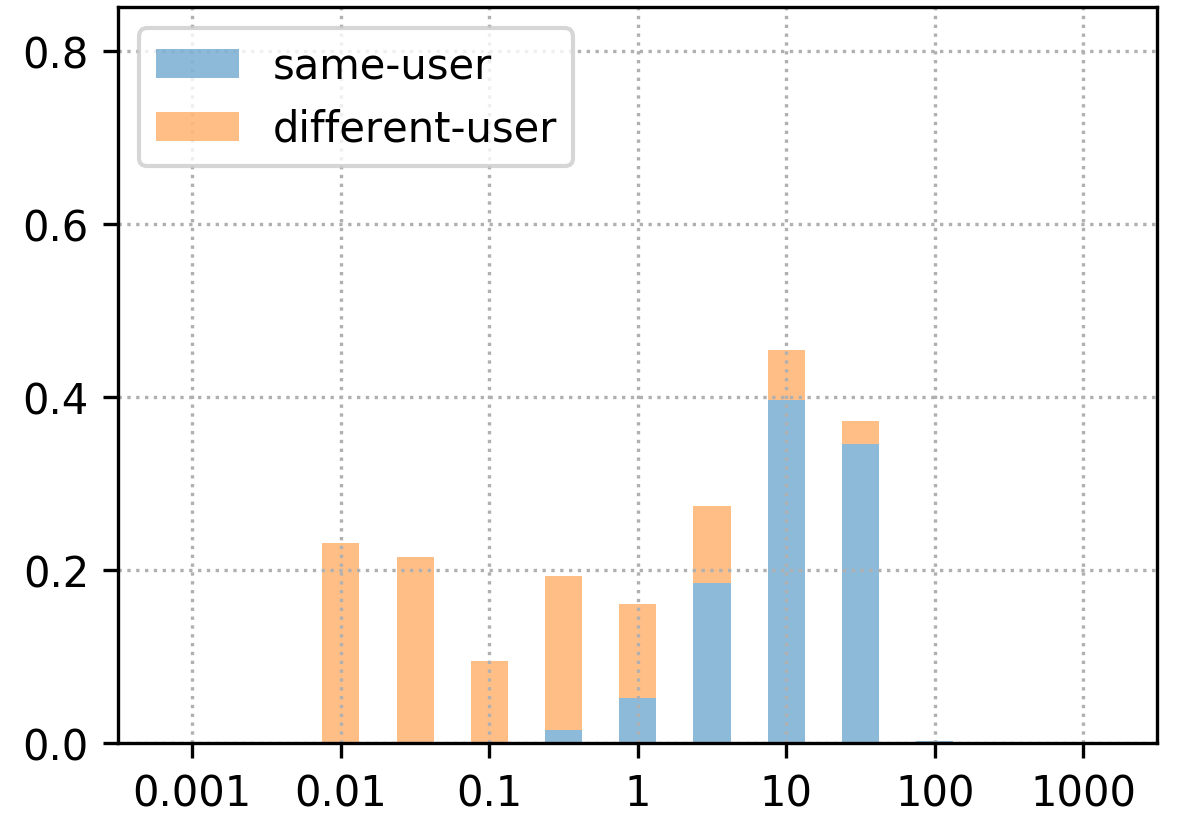}
\hspace*{5mm}log(LR)\hspace*{45mm}log(LR) \\
\caption{
Distribution of $\textrm{log}_{10}$ LR values for (blue) {\em same-user} and (orange) {\em different-user} track pairs, for (left) test and (right) validation data.
}
\label{fig:LR_distribution}
\end{figure}

Over the hundred train/test splits, our method achieves a mean log likelihood ratio cost ($C_{llr}$) of 0.47 (standard deviation 0.07).
On the real user validation data, the $C_{llr}$ is 0.47 (standard deviation 0.05).
Figure \ref{fig:LR_distribution} shows the distribution of LRs under both hypotheses, for both the test and validation data.
No indication of lack of calibration is visible, with strong misleading evidence rare.
The method seems better calibrated when applied to the test data, which is expected as these data are drawn from the same distribution as the train data.
For both test and validation data, average minimum and maximum LRs found are 1/78 and 57, indicating that the strength of evidence is relatively small, particular under $H_{su}$.

In general, simpler models perform worse (Figure \ref{fig:classifier_cllr}), with both the binning approach and the dislocations consistently giving higher $C_{llr}$ values on both the test and validation data.
However, the simple approach of looking at the mean distance performs as well as the baseline method on the test data ($C_{llr}=0.42$), and slightly better on the validation data ($C_{llr}=0.42$).

Performance of the more sophisticated models is mixed (Figure \ref{fig:classifier_cllr}).
In particular gradient boosting gives better results on the test data, but on the validation data the performance increase is small.
The support vector machine (SVM) and random forest perform worse on both datasets.
Note that we did not explore further hyperparameter tuning, which might increase model performance, at an increased risk of overfitting.

Most calibration methods perform similarly (Figure \ref{fig:calibrator_cllr}).
Changing the bandwith on the KDE or using isotonic regression (i.e. the PAV algorithm) yields very similar $C_{llr}$ values for both test and validation data.
Performing no calibration or using Gaussians yields worse $C_{llr}$ values on both datasets.

For both the test and validation data, there is an optimal fraction of training data to use for calibration (Figure \ref{fig:calibration_fraction_cllr}).
As expected, for both datasets using few data points for calibration leads to well trained models, i.e.\ low $C_{llr,min}$, but gives high $C_{llr}$ due to poor calibration.
Using few data points for model fitting leads to a high $C_{llr, min}$ and a high $C_{llr}$.
This is visually represented in Figure \ref{fig:calibration_fraction_cllr} by the blue bars increasing in height from left to right
and the orange bars decreasing, leading to a lowest combined height in the middle.

Very similar or slightly worse results are obtained when removing features (Figure \ref{fig:feature_columns_cllr}).
As expected, using the time difference distribution as sole feature leads to $C_{llr}=1$.
This is because the sampling leads to independent event times for all pairs of phones, meaning the time differences by themselves yield no information.
Interestingly, performance is similar to baseline for those combinations of features that include distance, and worse for those that do not include distance. 

Performance increases when more data points are available per track pair, as expected (Figure \ref{fig:number_of_days_cllr}). 
The performance drop we see when increasing the data per track pair from three to four consecutive days may well be due to the associated drop in number of track pairs, with only 28 samples per hypothesis available for calibration and 13 for testing.
Interestingly, this same pattern can be seen in attenuated form when we lower the sampling rate (Figure \ref{fig:events_per_hour_cllr}), indicating that the improved performance stems from more than an increase in data points. 
Thus, it seems that having larger time differences for switches increases the information content.

\newcommand{\figurepair}[3]{
\begin{figure}
\centering\footnotesize
~~reference phones\hspace*{.32\textwidth}user phones
\rotatebox{90}{\hspace*{25mm}$C_{llr}$}~~%
\includegraphics[width=0.49\textwidth]{plots/test_#1.png}%
\includegraphics[width=0.49\textwidth]{plots/val_#1.png}
#3
\caption{#2}
\label{fig:#1}
\end{figure}
}

\figurepair{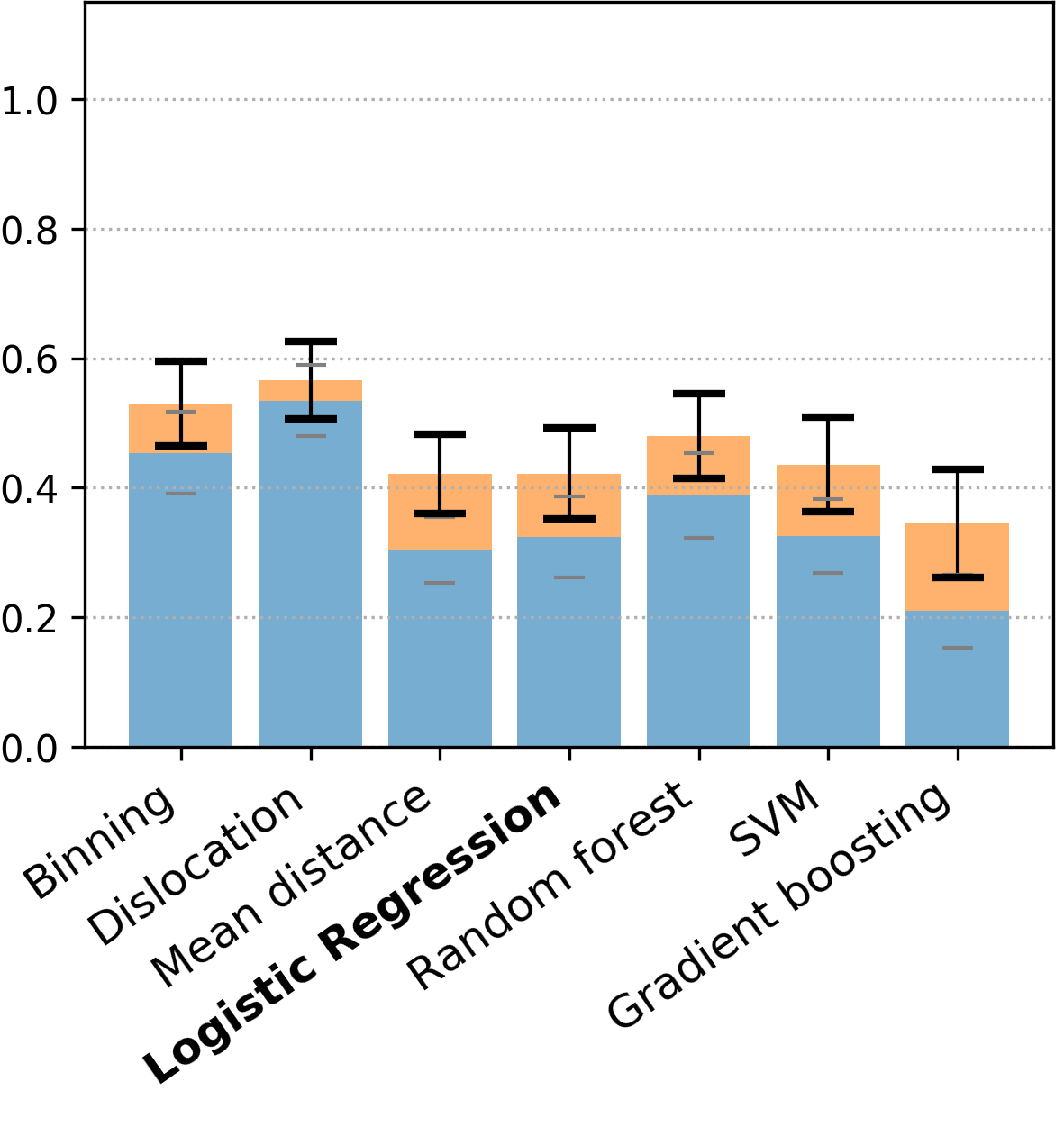}{
Average log likelihood ratio cost ($C_{llr}$) and standard deviation for the 100 runs, for a variety of methods explored, for (left) the test data and (right) the validation data.
Baseline method is indicated in bold.
The blue bar indicates the average $C_{llr,min}$.
Thus, the blue area is an indication of the discriminating power of the method, the orange area of its calibration.
The baseline method (in bold) uses a two-step logistic regression approach.
The first three methods are simpler, directly classifying track pairs by aggregating the switches' features.
The random forest, support vector machine (SVM) and gradient boosting use the same two-step approach, but a different mathematical model.
See section \ref{classifiers} for details.
}{Classifier}

\figurepair{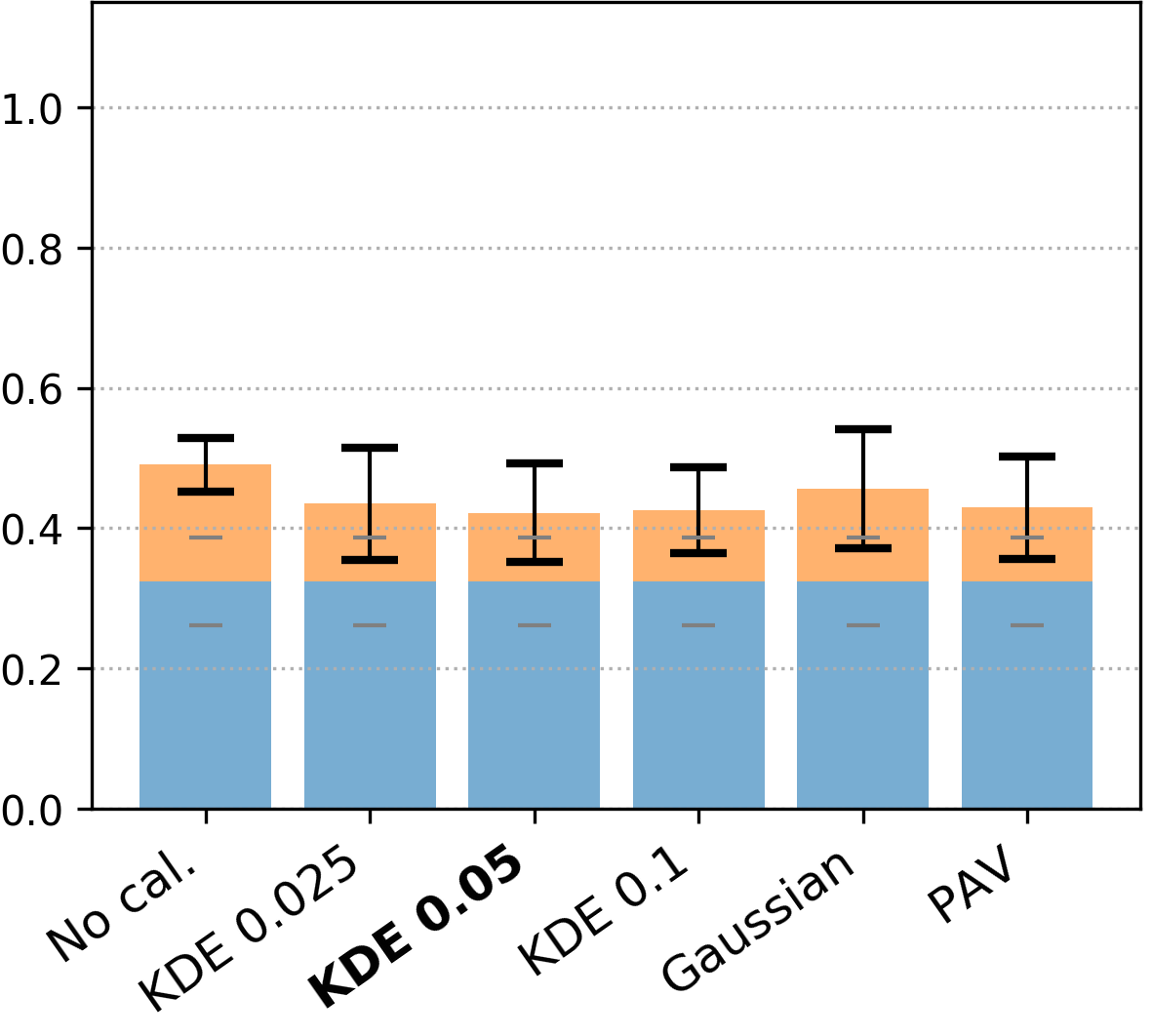}{
Average log likelihood ratio cost ($C_{llr}$) and standard deviation for the 100 runs, for a variety of calibration methods explored.
Baseline method is indicated in bold.
The blue bar indicates the average $C_{llr,min}$.
Thus, the blue area is an indication of the discriminating power of the method, the orange area of its calibration.
As expected, the blue area is the same for all methods.
Performance is similar for the different calibration methods, surprisingly even when no calibration is applied.
The exception is the Gaussian method, which performs consistently poor.
}{Calibration method}

\figurepair{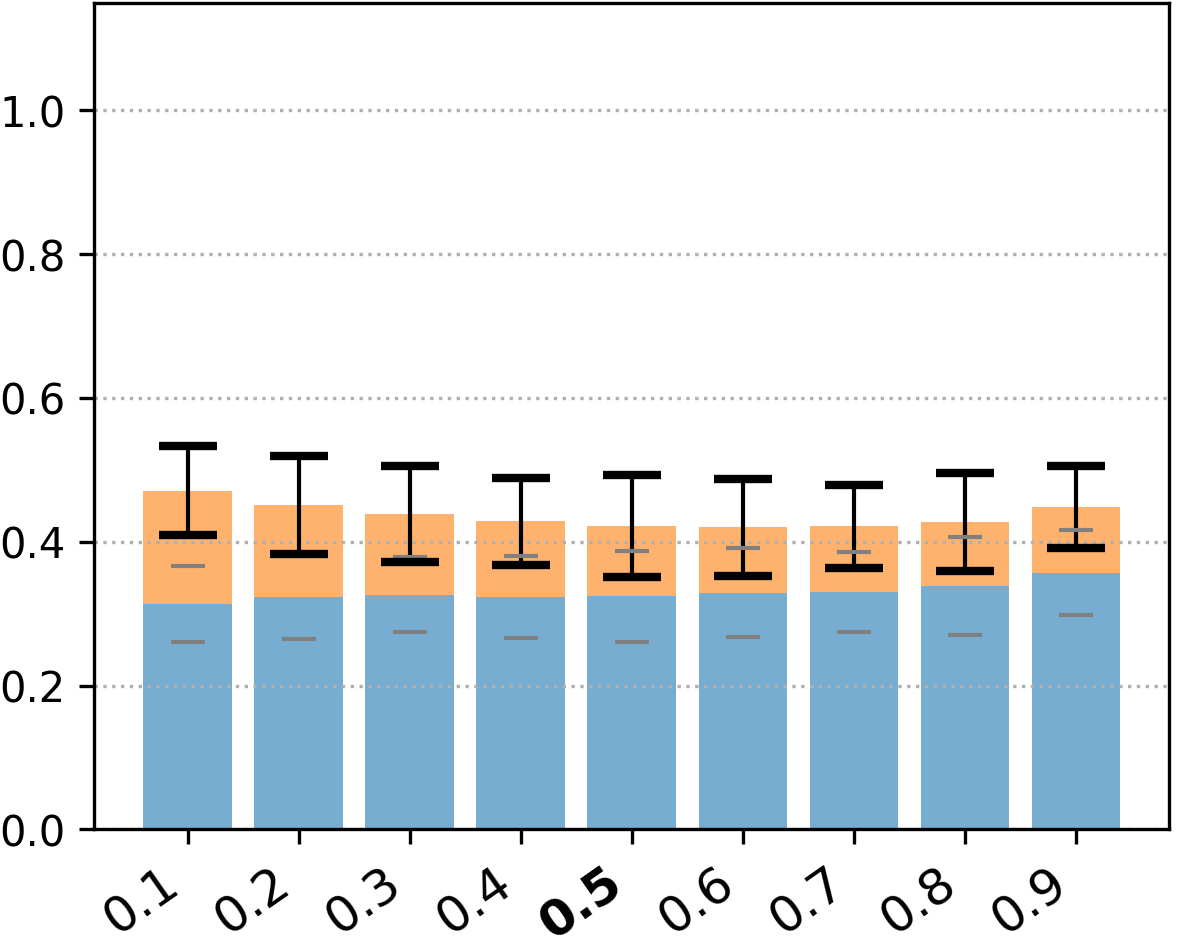}{
Average log likelihood ratio cost ($C_{llr}$) and standard deviation for the 100 runs, for the fraction of training data used for calibration varied from 0.1 to 0.9.
Baseline method is indicated in bold.
The blue bar indicates the average $C_{llr,min}$.
}{Calibration fraction}

\figurepair{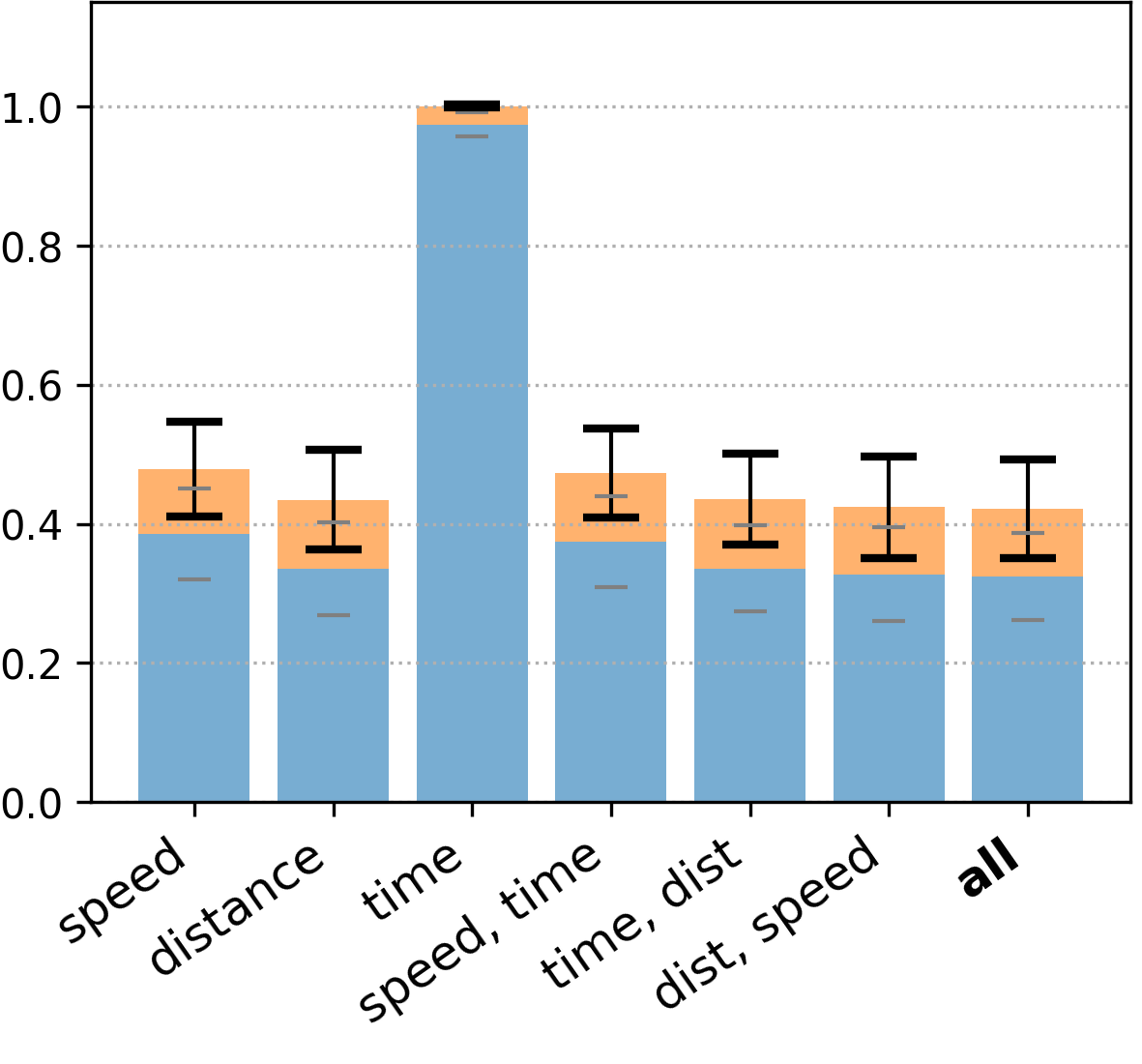}{
Average log likelihood ratio cost ($C_{llr}$) and standard deviation for the 100 runs, for different combinations of features used.
Baseline method is indicated in bold.
The blue bar indicates the average $C_{llr,min}$.
The baseline method uses the time difference, distance and associated speed for any switch.
Here results are given when a subset of those are used, or when a fourth feature, diffence in direction, is used.
}{Features used}

\figurepair{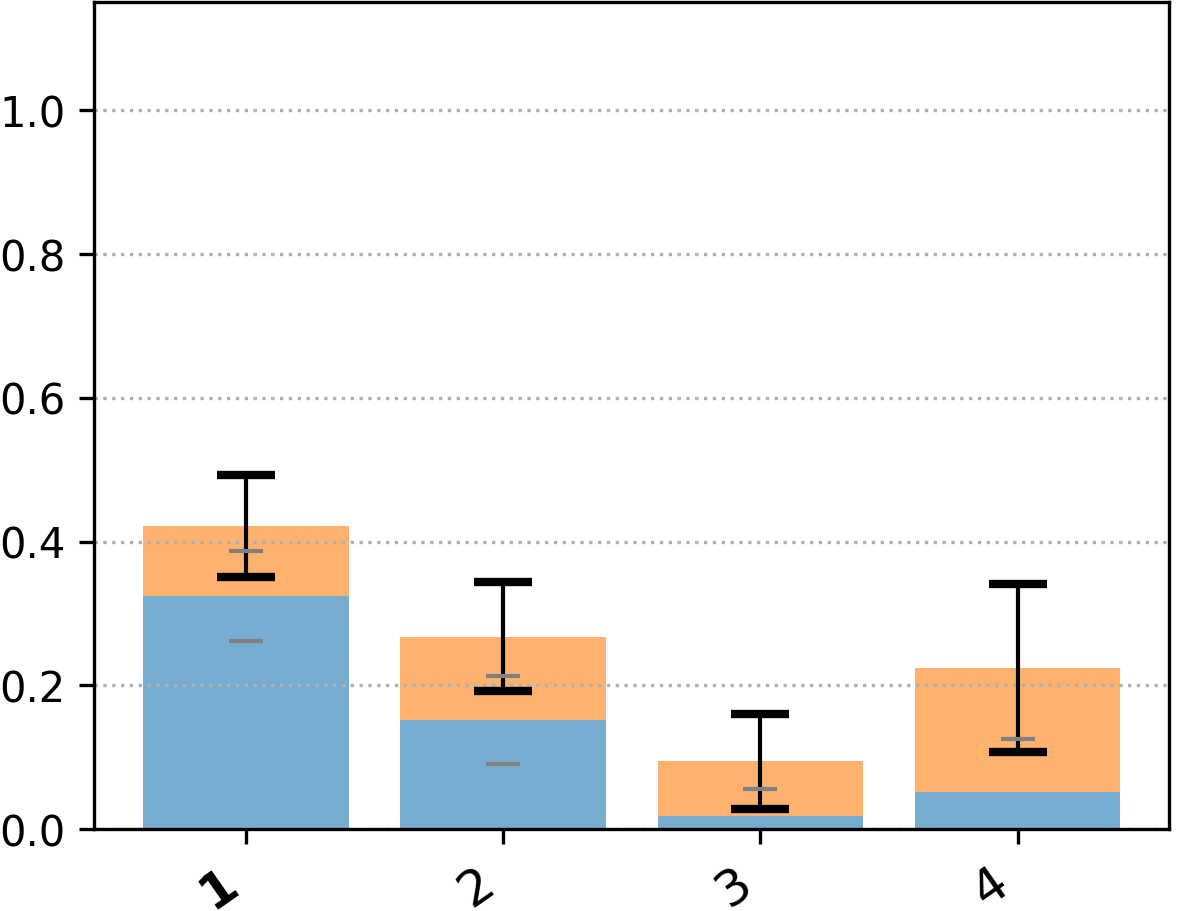}{
Average log likelihood ratio cost ($C_{llr}$) and standard deviation for the 100 runs, for a different number of days of data. 
Baseline method is indicated in bold.
In the validation data, the maximum number of consecutive days of data for {\em same-user} pairs is two.
}{Length of tracks (days)}

\figurepair{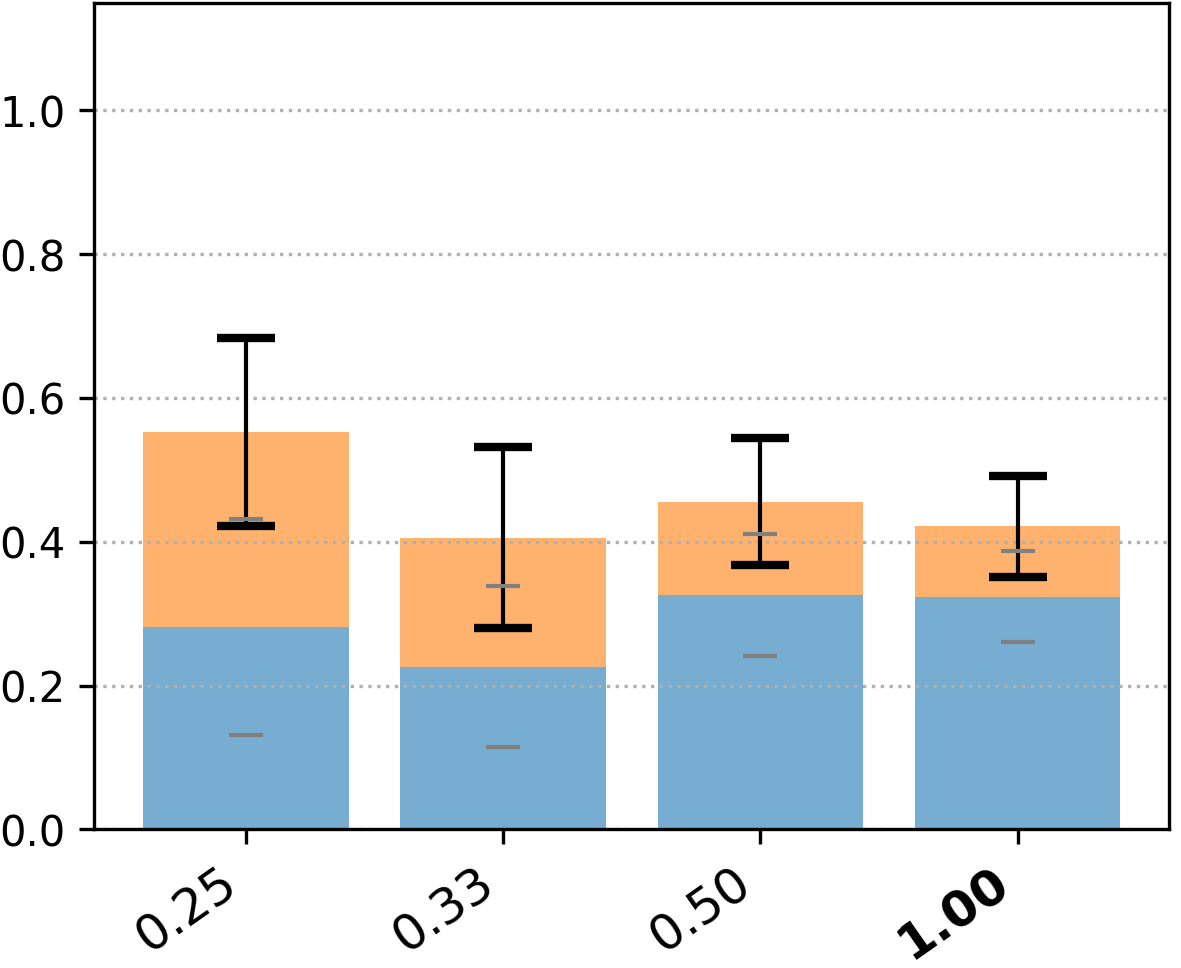}{
Average log likelihood ratio cost ($C_{llr}$) and standard deviation for the 100 runs, for different sampling rates. 
Baseline method is indicated in bold.
For each sampling rate, the number of days is adjusted so the expected number of samples remains the same (e.g. rate 0.5 is taken over 2 days). In the validation data, the maximum consecutive days of data for {\em same-user} pairs is two.
}{Events per hour}

\section{Discussion}
We presented a novel method for evaluating the strength of evidence from call detail records that any pair of phones were carried by the same person.
The method produces a score for pairs of registrations, which lead to a score for any pair of phones, for a given period. 
A calibration step follows, in which the score is converted into a likelihood ratio (LR) between both hypotheses.
Using data from experiments and real phone usage we assessed performance of the method. We further assessed the impact of a range of model or data changes.

Data is key to both the fitting of any statistical model and the validation of any LR method. 
One of the main contributions of this paper is the generation of a relatively large dataset of experimental data, and an additional validation on real user data.
However, these data have some clear limitations.

First, the phone users in these data are from a different population than the population of interest for criminal investigations. 
Most of our users show a somewhat predictive pattern of home-work travel which may not translate well to mobility patterns of potential suspects.
Forensic experts will have to consider for any given case whether this mismatch is too large to make the model applicable.
In our experience of criminal cases, we have mainly encountered suspects' travel patterns that were more erratic, making the model presented here conservative (in the sense that generated LRs were likely too close to 1).

Second, the experimental data is not representative of real user data, as evidenced by the lower performance on the validation data than on the test data.
A main difference may be the timing of events.
The homogeneous Poisson process used for sampling the experimental data is a simplistic model for phone usage.
In real data, we may expect inhomogeneities, and dependence between timing of events for a phone, and between phones of the same user.
In fact, we can see this dependence in the many small time differences for the real user data as compared to the experiment data (Figure \ref{fig:data}). 
One improvement on the current method may thus be to more accurately model typical behavior.
This could have the beneficial side effect of strengthening the method as more information (dependence of timing of events) can then be used.
Alternatively, in a criminal case the timings of events could be conditioned on the exact timings of the disputed phone pair.
This has the advantage of not having to make assumptions on phone usage \cite{vaneijk13a}, and staying closer to the question at hand.

Another difference between real user data and experimental data is that the experimental phones are guaranteed to be together in one package at all times, whereas the real users may at times leave one phone behind, during which time location registrations can still occur.
This adds noise to the user data which may hurt validation performance.

\subsection{Future work}

There is scope to improve the method by using more complex mathematical models, as evidenced by the better performance on the test data of gradient boosting (Figure \ref{fig:classifier_cllr}).
For example, removing features other than distance yields similar results with our method, even though a large distance in a large time frame is much less indicative of {\em different-user} pairs than the same large distance in a relatively short timeframe (this is roughly the idea behind dislocations).
This phenomenon is visible in figure \ref{fig:data} by the diagonal line in the {\em same-user} pairs that is absent in the {\em different-user} pairs.
Of course, our linear logistic regression classifier cannot model this dependence, as evidenced by the similar performance on the test data of using only the feature distance (Figure \ref{fig:feature_columns_cllr}), or even only mean distance (Figure \ref{fig:classifier_cllr}).
Our main reason to stick with logistic regression is the simpler interpretation of learned weights, and thus the higher explainability in court.
However, another good reason is that when there is a mismatch between the reference data and the data of interest (here: the validation data), the more complex models that yield relatively high LRs on the test data may yield too high LRs on the data of interest. 
We can see this in the relatively high $C_{llr}$ of gradient boosting on the validation set.
Interestingly, its $C_{llr,min}$ is still low, indicating that the discriminative power may still be higher, but the calibration has been overconfident.
When deciding which fraction of training data to devote to calibration, Figure \ref{fig:calibration_fraction_cllr} shows that there is an optimum around 50 percent for both the reference phones and the user phones.
Less data to calibration causes increased calibration loss; less data to model construction causes decreased discriminative power.

The method we have presented could be improved by considering typicality as well as similarity.
Generally, LR systems can consider two types of properties of a pair of measurement sets to compare: similarity and typicality.
For example, a system that computes LRs based on body length estimations could evaluate both the absolute difference between the estimates (similarity) and how rare these lengths are in the population (typicality). 
A system that only looks at the former will give the same LR for 180 cm vs 181 cm as for 230 cm vs 231 cm, even though intuitively it is clear the latter is much less likely to result by chance. 
The features our method looks at are examples of the similarity of two tracks.
Relevant information about typicality, e.g. the amount of movement of the phones, is not used.
Thus, currently two phones staying close to each other will get a similar LR regardless of whether they are traveling around the country or staying in one place.
We have partially remedied this issue here by discarding tracks with no or little movement.
A useful extension to the method would of course be to capture this typicality, and adjust the LRs accordingly.
Alternatively, in case work, a good approach may be to only take tracks from the reference data that are similar to the tracks from the disputed pair.

\subsection{Practical use}

The presented method has wider applicability than establishing co-usage of two phones. 
In essence, the method addresses the question how well the similarity between two sets of cell data records fits with a scenario where the associated SIM cards were traveling jointly or independently.
Thus, the method might be applicable to questions concerning the possible co-location of two persons, each with their own phone. 
Likewise, as many high-end cars communicate using the same technology, questions concerning co-location of a person and a car may be answerable. 
We should note that although this makes the method more applicable, it also means that due thought to such scenarios should be given when formulating hypotheses. 
For example, a defense hypothesis of two phones being carried by separate persons traveling together (rather than independently) would yield data patterns very similar to the {\em same user} hypothesis.

It is not the case that the proposed method is applicable or relevant for any court case. 
When considering or evaluating the use of this method in criminal court cases, the prosecutor, forensic expert and defense lawyer should consider the following.

First, is there enough data in the disputed tracks?
This questions does not affect applicability but efficiency of resource usage. 
Although we have shown the method works well even with only a day of data, if the day of data only comprises one or two data points for one of the phones, a comparison is likely to yield an LR close to 1.
Producing a full forensic report may thus not be cost-effective.

Second, are the relevant hypotheses consistent with the method and its reference data?
In particular, what is the relation between the phone users under the defense hypothesis?
Are they completely unrelated persons in 1) the same country or 2) the same city, or may they be 3) vaguely or 4) intimately known to each other?
These four possibilities would yield an LR sequentially close to 1 for the same data, as the expected phone pair characteristics under the defense hypothesis are increasingly close to the {\em same-user} pair characteristics of the prosecutor hypothesis.
The experiments described in this paper fit best with the third possibility, as the test subjects are colleagues.

Third, do the available reference data adequately represent the disputed pair?
This is the hardest consideration, but possibly the most important one.
It seems very likely that there is a mismatch in spatiotemporal patterns between the test population and those of suspects and criminals, e.g. in times and amount of travel and variety of destinations.
Does this mismatch prohibit application of the method?
In casework we have argued that the test population has a higher predictability of travel, thus making it harder for a classifier to distinguish between {\em same-user} and {\em different-user} than when travel would be more erratic.
An LR system trained on this population would thus yield LRs closer to 1 than when it had been trained on a more representative erratically traveling population.
This makes it conservative for case work.
However, this question should be reviewed for every case.

\subsection{Conclusion}
We present a forensic method to quantify the notion that two phone tracks look `more similar than expected by chance'.
The method may be considered for use in court, but its applicability should be reviewed on a case by case basis.

\ifcsname anoniem\endcsname%
\else
\section{Acknowledgements}
We are grateful to anyone who supported this project and helped getting high quality data,
including volunteers who have shared their cell phone data, or participated in the field experiments.
We further thank Cor Veenman, Jeanette Leegwater and the reviewers for helpful suggestions and critical reading.
\fi

\bibliographystyle{elsarticle-num}
\bibliography{doc}
\end{document}

%% file: revision.tex
\newcommand{\revision}{git:c6d2833}